\documentclass[11pt,onecolumn,draftclsnofoot]{IEEEtran}
\usepackage{verbatim,amsfonts,graphicx,amsmath,amsbsy,amssymb,citesort,epsfig,url}

\newtheorem{THEO}{Theorem}

\def \b {{\mu}}
\def \xhatn {{\widehat{x^n}}}
\def \zhatn {{\widehat{z^n}}}
\def \xn {{x^n}}
\def \yn {{y^n}}
\def \ynr {{y^n_r}}
\def \zn {{z^n}}
\def \znr {{z^n_r}}
\def \C{{\cal{C}}}
\def \I {{\cal{I}}}
\def \X {{\cal{X}}}
\def \Y {{\cal{Y}}}
\def \Z {{\cal{Z}}}
\def \Ze {{{\cal{Z}}_e}}
\def \overC {{\overline{{\cal{C}}}}}
\def \overY {{\overline{{\cal{Y}}}}}
\def \overcj {{\overline{c_j}}}
\def \overcij {{\overline{c_{ij}}}}

\newlength{\algorithmwidth}
\begin{document}

\title{\vspace*{-15mm}
An MCMC Approach\\ to Universal Lossy Compression\\ of Analog Sources\footnote{
Much of the research was performed when the first author  was with 
the Electrical Engineering Department at the Technion, Israel.
Subsets of the work appeared in~\cite{BaronWeissmanDCC2010}.}
}
{\small \author{
Dror Baron \\
Department of Electrical and Computer Engineering \\
North Carolina State University; Raleigh, NC \\
Email: barondror@ncsu.edu \\
\and
Tsachy Weissman\\
Department of Electrical Engineering \\
Stanford University; Stanford, CA \\
Email: tsachy@stanford.edu
} }
\maketitle

\begin{abstract}
Motivated by the Markov chain Monte Carlo (MCMC) approach to the compression
of discrete sources developed by Jalali and Weissman, 
we propose a lossy compression algorithm for analog
sources that relies on a finite reproduction alphabet, which
grows with the input length. The algorithm achieves, in an appropriate
asymptotic sense, the optimum Shannon theoretic tradeoff between 
rate and distortion, universally for stationary ergodic
continuous amplitude sources.
We further  propose an MCMC-based algorithm that resorts to a reduced
reproduction alphabet when such reduction does not prevent
achieving the Shannon limit. The latter algorithm is 
advantageous due to its reduced complexity and improved rates of 
convergence when employed on sources with a finite and small
optimum reproduction alphabet.
\end{abstract}


\section{Introduction}

Lossy compression of analog sources is a pillar of
modern communication systems. Despite numerous applications
such as image compression~\cite{xiong97sf,Lopresto1997}, video
compression~\cite{Wiegand2003}, and speech coding~\cite{Makhoul85,buzo1980speech,sabin1984product},
there is a significant gap between theory and practice.

\subsection{Entropy coding}

\begin{figure*}[t]
\begin{center}
\includegraphics[width=120mm]{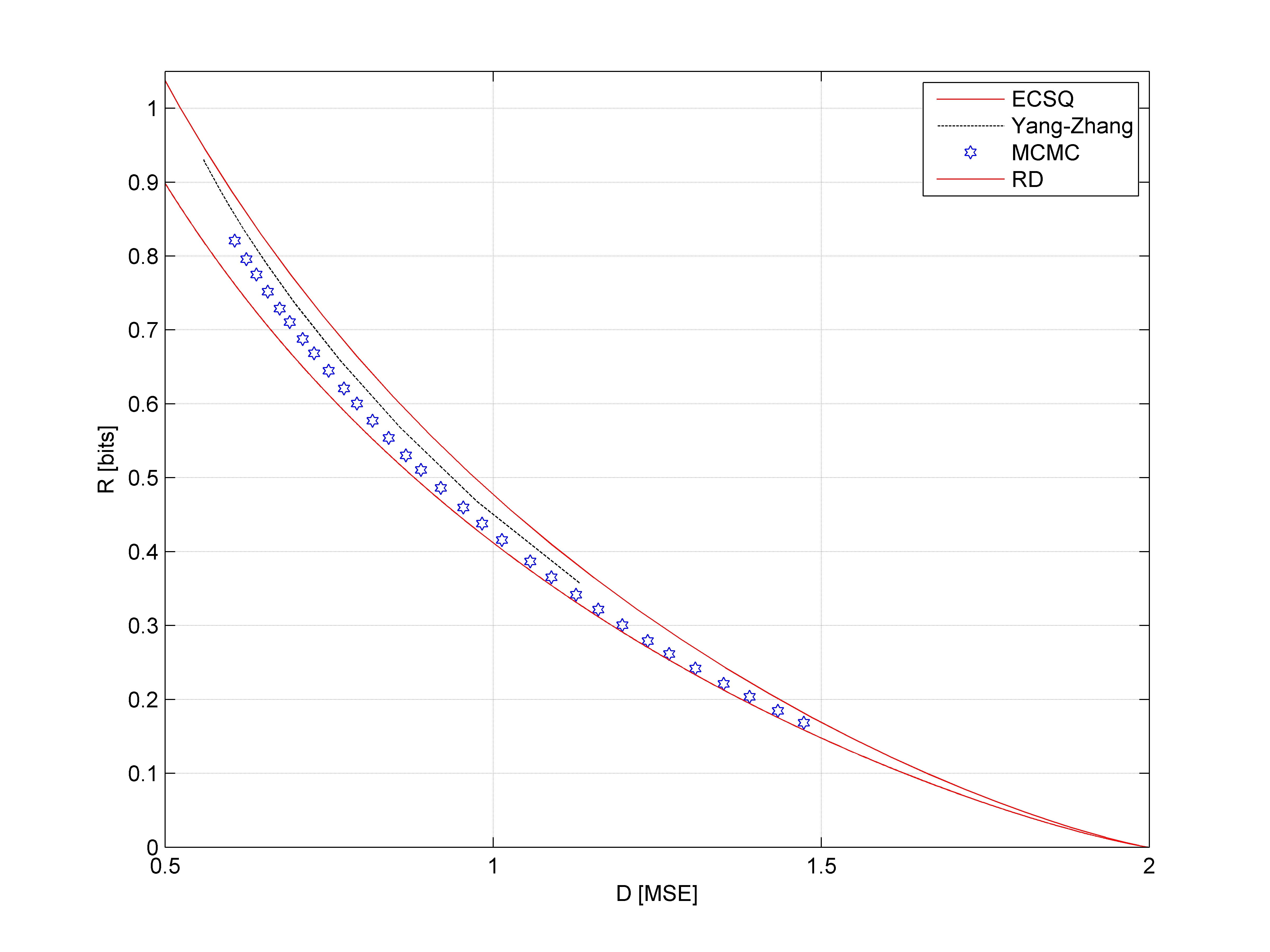}
\end{center}
\vspace*{-8mm}
\caption{
{\small\sl {\bf Laplace source}:\
Comparison of entropy coding (ECSQ),
results by Yang and Zhang~\cite{Yang1999},
average rate and distortion of
Algorithm~2 (MCMC) over 10 simulations,
and the RD function.
($n=1.5\cdot 10^4$,  $|\Z|=9$, $r=50$,
$k\approx\frac{1}{2}\log_{|\Z|}(n)$.)}
\label{fig:MCMC_Laplace}
}
\vspace*{-5mm}
\end{figure*}

Many practical lossy compression algorithms employ {\em entropy coding}, 
where {\em scalar quantization} is followed by lossless compression (ECSQ). 
ECSQ has motivated much work into optimization of scalar quantizers~\cite{FarvardinModestino84,Lloyd82,Max60},
whereas the translation to bits can use Huffman~\cite{Huffman52}
or arithmetic~\cite{RissanenLangdon1981,Cover91} codes. 
Despite the simplicity and elegance of ECSQ,
even for independent and identically distributed (iid) sources
the {\em rate distortion} (RD) function~\cite{Berger71,Cover91},
which characterizes the fundamental limit for lossy compression, 
suggests that ECSQ-based coding can be highly suboptimal 
(cf. Figure~\ref{fig:MCMC_Laplace} for an example).
For non-iid sources, ECSQ may compare even less favorably
with the fundamental RD limit.

In order to bridge part of the gap between ECSQ and the RD function,
{\em vector quantization} (VQ) converts an entire vector to a 
codeword~\cite{chou1989entropy,sabin1984product,riskin1991greedy}, 
in contrast to scalar quantization, which compresses individual scalar input elements.
VQ provides a better trade-off between rate and distortion
as the vector dimension increases,
but increased complexity is required \cite{GershoGray1993}.
The significant computation required by VQ
necessitates developing computationally feasible alternatives.

\subsection{Related work}

For {\em finite alphabet} sources, recent advances
have demonstrated that the RD limit can be approached
asymptotically~\cite{GioranKontoyiannis2009,GVW2008b,Kontoyiannis1999} by
{\em partitioning an input into sub-blocks}, where a Shannon-style
codebook~\cite{Cover91,Berger71} is applied to each sub-block. 
Some of these schemes can compress {\em universally} without knowing the source
statistics beforehand, but it is challenging to generate a codebook distribution
whose statistics differ from those of the input statistics~\cite{Zamir2001}.

Lossy compression over a finite alphabet can also be performed by
{\em directly mapping the entire input to an output sequence} 
while accounting for the trade-off
between the compressibility of the output and the distortion between the input and
output sequences. This optimization can be deterministic~\cite{Yang1997} or
stochastic~\cite{Jalali2008} in nature. Directly mapping to the
output sequence effectively quantizes the entire input -- a long sequence -- into
a large output codebook, and achieves the RD limit for
stationary ergodic finite alphabet sources universally.
Another promising recent approach to (non-universal) lossy compression relies on
algebraic codes~\cite{Hussami2009}.

For {\em analog sources}, less progress has been made in developing
theoretically-justified compression algorithms. Some results have been derived
specifically for the high-rate regime, where the Shannon lower bound is asymptotically
tight~\cite{LinderZamir94}. In particular, in the limit of low distortions the
RD limit has been
characterized for mixtures of {\em probability distribution functions} (pdf's)
where one distribution is discrete and the other continuous~\cite{GLZ99,RosenthalBinia88}.
For example, the sparse Gaussian source is a mixture pdf; bounds on its RD function have
been provided~\cite{WeidmannV2008,WeidmannV99,CWO2002,ChangSparseRD2009}.

Despite the theoretical insights in the high-rate regime, compression of analog sources
at {\em low-to-medium rates} is of interest in many
applications~\cite{xiong97sf,Lopresto1997,Wiegand2003,Makhoul85}.
There do exist special input pdf's for which
entropy coding approaches the RD function~\cite{MarcoNeuhoff2006} in the low-rate limit,
but the low-rate regime is challenging in general.
We aspire to develop results of general applicability and not be limited to specific pdf's
with fortuitous properties.

\subsection{Contributions}

The crux of our approach to the compression of analog sources is to quantize 
to discrete
reproduction levels, and then apply a compression algorithm 
similar to that of
Jalali and Weissman~\cite{Jalali2008}, which uses
the stochastic optimization approach of
{\em Markov chain Monte Carlo} (MCMC) simulated annealing,
as championed in~\cite{Geman1984}.
A careful choice of the set of reproduction levels,
growing appropriately with input length both in size and in resolution,
achieves the RD function despite
the analog nature of the source.
A somewhat similar approach was suggested by Yang et al.~\cite{Yang1997,Yang1999}
using deterministic optimization techniques.
Note, however, that Yang and Zhang~\cite{Yang1999} require
availability of a training sequence, and so their algorithm is not universal.
Although it is possible to apply deterministic optimization in
a universal setting by partitioning the input into blocks, this approach
results in a performance loss of 0.2--0.3 dB~\cite{Yang1999}.

Our first contribution is a lossy compression algorithm for analog
sources that relies on a {\em data-independent reproduction alphabet} that
grows with the input length. 
This algorithm asymptotically achieves the
RD function universally for stationary ergodic continuous amplitude
sources. However, the reproduction alphabet grows with the input length,
slowing down the convergence to
the RD function, and is thus an impediment in practice.

To address this issue,  we next propose an MCMC-based algorithm 
that uses an
{\em adaptive reproduction alphabet}. The ground-breaking work by Rose on
the discrete nature of the reproduction alphabet for iid sources
when the Shannon
lower bound is not tight~\cite{Rose94} suggests that, for most sources
of practical interest, restriction of the reconstruction to
 a rather small alphabet does not stand in the way of attaining the
fundamental compression limits.
Indeed, at low rates even a binary reproduction alphabet is often 
optimal~\cite{MarcoNeuhoff2006}.
When employed on such sources, our latter algorithm zeroes in on the finite
reproduction alphabet, and thus enjoys rates of convergence commensurate 
with the finite-alphabet setting.

In order to render this adaptive algorithm computationally feasible,
we develop a method to update the optimal reproduction levels
rapidly. Utilizing this computational feature,
our adaptive algorithm provides faster computation, 
achieves the RD function universally, and in some cases the smaller reproduction
alphabet accelerates convergence to the RD function. Consequently, the adaptive algorithm
is more suitable in practice.
We emphasize that our algorithms are both universal, requiring no knowledge of
the source statistics.

The remainder of the paper is organized as follows.
We provide background information in Section~\ref{sec:background}.
Our first, brute force algorithm is described in Section~\ref{sec:naive},
followed by the adaptive reproduction alphabet algorithm
in Section~\ref{sec:adaptive}.
Numerical results are reported in Section~\ref{sec:numerical}.
We complete the paper with a discussion in Section~\ref{sec:discussion}.
Proofs appear in appendices, in order to make the main portion of the
manuscript easily accessible.

\section{Background}
\label{sec:background}

\subsection{Notation}

Consider a stationary ergodic source $X=\{X_i, i\geq 1\}$ 
with real-valued components.
We process a length-$n$ input $\xn=x_1x_2\ldots x_n$, which is an individual
realization of the random vector $X^n$. The input $\xn$ is compressed
using an {\em encoder} $e:\X$$^n \rightarrow \{0,1\}$$^+$
that maps $\xn$ to a finite output string $e(\xn)$.
The {\em decoder} $d: \{0,1\}^+\rightarrow \Y$$^n$ maps the bit string
back to a length-$n$ output $\yn$ over the reproduction alphabet
$\Y$, which may be a continuous or discrete subset of the real line.
The output $\yn$ is the lossy approximation of $\xn$.

We assess the performance of an encoder-decoder pair 
relative to the trade-off between rate and distortion~\cite{Cover91,Berger71}.
The {\em rate} of such a pair is defined as $R=E[\frac{1}{n}|e(X^n)|]$,
the expected number of bits per description of a source
symbol,
where $|\cdot|$ denotes length, size, or cardinality, and $E[\cdot]$ is expectation.
The {\em distortion} $D=E[d_n(X^n,\yn)]$ quantifies the expected per-symbol distortion,
\begin{equation}
\label{eq:def:d_n}
d_n(\xn,\yn) \triangleq \frac{1}{n} \sum_{i=1}^n d(x_i,y_i),
\end{equation}
where $d:\mathbb{R} \times\mathbb{R} \rightarrow\mathbb{R}^+$ 
measures the distortion. For concreteness in what follows, we assume the
distortion is the square of the error $d(x_i,y_i)=(x_i-y_i)^2$, but 
our approach readily carries over to accommodate general distortion measures.

\subsection{Lossy compression using MCMC}
\label{subsec:MCMC}

We describe a variant of the scheme in~\cite{Jalali2008}
that compresses an input $x^n$ to an output $y^n$ over a finite 
alphabet $\Y \subseteq \mathbb{R}$.
This algorithm will later be employed as the main building block 
for compressing an analog source.
The encoder approximates $\xn$ by $\yn$, which is compressed using the
{\em context tree weighting} (CTW) universal lossless compression algorithm.\footnote{
We prefer CTW~\cite{Willems1995CTW}, because for context tree sources it has
lower redundancy than Lempel-Ziv based schemes~\cite{LZ78} or adaptive arithmetic
coding based on full-tree Markov models~\cite{Yang1999}.
}
The approximation $\yn$ is chosen to provide a good trade-off 
between the coding length required for $\yn$ and the distortion with respect to $\xn$.
The decoding procedure is straightforward; the output bits are passed through the
CTW decompressor to retrieve $\yn$.

Denote the empirical symbol counts by $m_k(\yn,u^k)[a]$, i.e., 
\[
m_k(\yn,u^k)[a] \triangleq | \{ k< i \leq n: y_{i-k}^i=u^k a \} |,
\]
where $k$ is the context depth, $a\in\Y$, $u^k\in\Y^k$,
and $u^k a$ denotes concatenation of $u^k$ and $a$.
Define the $k$-depth conditional empirical entropy as
\begin{equation}
\label{eq:def:H}
H_k(\yn) \triangleq -\frac{1}{n} \sum_{a,u^k} m_k(\yn,u^k)[a]
\log\left( \frac{  m_k(\yn,u^k)[a] } { \sum_{a'} m_k(\yn,u^k)[a'] } \right),
\end{equation}
where $\log(\cdot)$ is the base-two logarithm, and we use the convention
wherein $0\log(0)=0$. For $k=o(\log(n))$, the 
difference between the CTW coding
length and the empirical conditional entropy is $o(1)$~\cite{Willems1995CTW}.
We  define the energy $\varepsilon(\yn)$ corresponding to $\yn$ by
\begin{equation}
\label{eq:MCMC_energy}
\varepsilon(\yn) \triangleq n[H_k(\yn)-\beta d_n(\xn,\yn)],
\end{equation}
where $\beta<0$ is the slope of the RD function at the point we want to attain.
The Boltzmann probability mass function (pmf) is
\begin{equation}
\label{eq:def_Boltzmann}
f_s(\yn) \triangleq \frac{1}{Z_s} \exp\{-s \varepsilon(\yn)\},
\end{equation}
where $s>0$ is inversely related to temperature in simulated
annealing~\cite{Geman1984}, and $Z_s$ is the normalization constant.

Ideally, our goal is to compute the globally minimum energy
solution $\xhatn$,
\begin{equation}
\label{eq:def:xhat}
\xhatn \triangleq
\arg\min_{w^n \in \Y^n } \varepsilon(w^n)
=\arg\min_{w^n \in \Y^n} [H_k(w^n)-\beta d_n(\xn,w^n)].
\end{equation}
Computation of $\xhatn$ involves an exhaustive search over
exponentially many sequences and is thus infeasible.
We use 
a stochastic {\em Markov chain Monte Carlo} (MCMC)
relaxation~\cite{Geman1984} 
to approximate the globally minimum solution,
in contrast to the deterministic approach of Yang et al.~\cite{Yang1997}.
We denote the resulting approximation by $\yn$.

To sample from the Boltzmann pmf (\ref{eq:def_Boltzmann}),
we examine all $n$ locations. For each location, we use a Gibbs sampler
to resample from the distribution of
$y_i$ conditioned on $y^{n\backslash i} \triangleq \{ y_n:\ n \neq i\}$
as induced by the joint pmf in (\ref{eq:def_Boltzmann}), 
readily computed to be
\begin{equation}
\label{eq:y_n_i}
f_s(y_i=a|y^{n\backslash i})
= \frac{1}
{ \sum_b \exp\left\{ -s\left[ n\Delta H_k(y^{i-1}by_{i+1}^n,a) -
\beta \Delta d(b,a,x_i) \right] \right\} },
\end{equation}
where $\Delta H_k(y^{i-1}by_{i+1}^n,a)$ is the change in $H_k(\yn)$
(\ref{eq:def:H}) when $y_i=a$ is replaced by $b$, and
$\Delta d(b,a,x_i)=d(b,x_i)-d(a,x_i) = (b-x_i)^2-(a-x_i)^2$ 
is the change in distortion.
We refer to the resampling from a single location as an iteration,
and group the $n$ possible locations into super-iterations.\footnote{
We recommend an ordering where each super-iteration scans
a permutation of all $n$ locations of the input, because in this manner
each location is scanned fairly often. Other orderings
are possible, including a completely random order
as prescribed by Jalali and Weissman~\cite{Jalali2008}.}

During the simulated annealing, the inverse temperature $s$
is gradually increased, where in
super-iteration $t$ we use $s=O(\log(t))$~\cite{Geman1984,Jalali2008}.
As the number of iterations $t$ is increased, $\yn$ converges in distribution
to the set of minimal energy solutions, which includes $\xhatn$
(\ref{eq:def:xhat}), because large $s$ favors low-energy $\yn$.
Pseudo-code for our encoder appears in Algorithm~1 below.

\fbox{\parbox{\algorithmwidth}{
\textsc{\underline{Algorithm~1}: Lossy encoder with fixed reproduction alphabet} \\
\textsc{Input:} $\xn\in\mathbb{R}^n$, $\Y$, $\beta$, $c$, $r$ \\
\textsc{Output:} bit-stream \\
\textsc{Procedure:}
\begin{enumerate}
\vspace*{-3mm}
\item Initialize $y$ by quantizing $x$ with $\Y$
\vspace*{-3mm}
\item Initialize $m_k(\cdot,\cdot)$ using $y$
\vspace*{-3mm}
\item {\bf for} $t=1$ to $r$ {\bf do} // {\em super-iteration}
\vspace*{-3mm}
\item \hspace*{5mm} $s \leftarrow c\log(t)$ for some $c>0$
// {\em inverse temperature}
\label{algo:const:temperature}
\vspace*{-3mm}
\item \hspace*{5mm} Draw permutation of numbers $\{1,\ldots,n\}$ at random
\vspace*{-3mm}
\item \hspace*{5mm} {\bf for} $t'=1$ to $n$ {\bf do} 
\vspace*{-3mm}
\item \hspace*{10mm} Let~$i$ be component $t'$ in permutation
\label{algo:const:inner1}
\vspace*{-3mm}
\item \hspace*{10mm} Generate new $y_i$ using $f_s(y_i=\cdot|y^{n\backslash i})$ given in (\ref{eq:y_n_i})
// {\em Gibbs sampling}
\vspace*{-3mm}
\item \hspace*{10mm} Update $m_k(\cdot,\cdot)[\cdot]$
\label{algo:const:inner2}
\vspace*{-3mm}
\item Apply CTW to $\yn$ // {\em compress outcome}
\end{enumerate}
}}

\section{Universal algorithm with data-independent reproduction alphabet}
\label{sec:naive}

Let us consider how Algorithm~1 can be
used to compress analog sources. 
We will see that choosing the
reproduction alphabet $\Y$ to be a finite subset of $\mathbb{R}$
(but growing with the input length $n$ in a data-independent way)
achieves the RD function.

Let us assume that the variance of source symbols
emitted by $X$ is finite, and consider the following data-independent 
reproduction alphabet,
\begin{equation}
\label{eq:def:overlineY}
\overY \triangleq \left\{
-\frac{\gamma^2}{\gamma},-\frac{\gamma^2-1}{\gamma},\ldots,
\frac{\gamma^2}{\gamma}
\right\}, \quad
\gamma=\lceil\log(n)\rceil,
\end{equation}
where  $\lceil\cdot\rceil$ denotes rounding up. 
In words, $\overY$ is a quantization of the interval 
$ [-\gamma,\gamma]$ to resolution $1/\sqrt{\gamma}$.
Other choices of $\overY$ also allow
to demonstrate various RD results; an examination of (\ref{eq:ap:th2_D1})
indicates that slower-growing $\gamma(n)$ also achieves the RD function.
The essential point is that $\overY$ quantizes a wider interval with
finer resolution as $n$ is increased, and its size increases sufficiently slowly
with $n$.

To prove achievability of
the RD function asymptotically, we first prove that a global optimization
(\ref{eq:def:xhat}) that determines $\xhatn$ followed by lossless compression
with CTW~\cite{Willems1995CTW} achieves the RD function.
Yang et al.~\cite{Yang1999,Yang1997} proved a similar result for
their deterministic algorithm while relying on a different reproduction
alphabet; our contribution is to prove achievability using the data-independent
reproduction alphabet $\overY$.

\begin{THEO}
\label{th:achieve_fixed}
Consider square error distortion (\ref{eq:def:d_n}),
let $X$ be a finite variance stationary and ergodic source with RD function
$R(X,D)$, and use the data-independent 
reproduction alphabet $\overY$ 
(\ref{eq:def:overlineY})
to approximate $\xn$ by the 
globally minimum energy solution $\xhatn$ (\ref{eq:def:xhat}). Then
the length of context tree weighting (CTW)~\cite{Willems1995CTW} 
applied to $\xhatn$ converges as follows,
\begin{equation}
\label{eq:th:achieve_fixed}
\lim_{n\rightarrow\infty}\sup
E\left[ \frac{1}{n}|CTW(\xhatn)|-\beta d_n(\xn,\xhatn)\right] \leq
\min_{D \geq 0} [R(X,D)-\beta D].
\end{equation}
\end{THEO}

Note that the $\lim\sup$ in (\ref{eq:th:achieve_fixed}) is actually a 
limit since the expectation on the left hand side is lower bounded by the right 
hand side for any scheme and any $n$, cf., 
e.g.,~\cite{Yang1997,Jalali2008}.
The detailed proof appears in~\ref{ap:achieve_fixed}, and we feature some
highlights here.
In order to prove achievability for the continuous alphabet source~$X$,
we construct a near-optimal codebook for a given input length $n$~\cite{Berger71}, 
and then quantize the components of every codeword in the codebook to $\overY$. 
As $n$ is increased,
$\overY$ quantizes a wider interval of values more finely.
The wider interval ensures that outlier source symbols have a vanishing effect on the
distortion, and finer quantization provides near-optimal distortion within
the interval. Therefore, we have achievability for the continuous amplitude
source~$X$ via the finite alphabet $\overY$.

Now consider running Algorithm~1 instead of the global energy minimization
(\ref{eq:def:xhat}) using the data-independent reproduction alphabet $\overY$.
The constant~$c$ used in Line~\ref{algo:const:temperature} of Algorithm~1
plays a crucial role. If $c$ is large, 
then the Boltzmann pmf (\ref{eq:def_Boltzmann}) favors
low-energy sequences too greedily, and the algorithm might get stuck in local minima. 
On the other hand, there exists a universal constant $c_1$ such that for $c<c_1$
we obtain universal performance. To understand why this happens, 
observe that Algorithm~1 optimizes over $|\overY|^n$ possible outputs.  
As long as $c<c_1$, there is 
a sufficiently large probability to transition between any two outputs,
and the algorithm cannot get bogged down in a local mimimum.
Therefore, in the limit of many iterations Algorithm~1 converges 
in distribution
to the set of minimal energy solutions, and we enjoy the same RD performance as
in Theorem~\ref{th:achieve_fixed}.
We refer the reader to Geman and Geman~\cite{Geman1984} 
for further discussions relating to the choice of $c_1$.
The proof appears in~\ref{ap:MCMC_fixed}.

\begin{THEO}
\label{th:MCMC_fixed}
Consider square error distortion (\ref{eq:def:d_n}),
let $X$ be a finite variance stationary and ergodic source with RD function
$R(X,D)$, and use  Algorithm~1 with the data-independent 
reproduction alphabet $\overY$ (\ref{eq:def:overlineY})
and sufficiently small $c<c_1$.
Let $\ynr$ be the MCMC approximation to $\xn$ after $r$ super-iterations. Then
the length of context tree weighting (CTW)~\cite{Willems1995CTW} 
applied to $\ynr$ converges as follows,
\[
\lim_{n\rightarrow\infty}
\lim_{r\rightarrow\infty}
E\left[ \frac{1}{n}|CTW(\ynr)|-\beta d_n(\xn,\ynr)\right]
\overset{n\rightarrow\infty}{\longrightarrow}
\min_{D \geq 0} [R(X,D)-\beta D].
\]
\end{THEO}

An important feature of the algorithm is that each iteration of
Lines~\ref{algo:const:inner1}--\ref{algo:const:inner2} requires
computation that is proportional to the context depth $k$
and alphabet size $|\overY|$, independent of $n$~\cite{Jalali2008}.
Because the alphabet grows slowly in $n$, the per-iteration computational
costs are modest. Each super-iteration contains $n$ iterations, and
so its computation is $O(nk|\overY|)=o(n\log^3(n))$.
Decoding is also fast. We first decompress CTW~\cite{Willems1995CTW},
and the finite alphabet is then mapped to our data-independent reproduction alphabet
$\overY$.

It is also noteworthy that our results could be modified to support other
distortion metrics. For example, if we used $\ell_p$ distortion, then
a technical condition $E[|X|^p]<\infty$ ensures
that outliers do not increase the distortion by much.

Although promising from a theoretical perspective, Algorithm~1 is of limited practical interest.
In order to approach the RD function closely, $\overY$
may need to be large,
which slows down the algorithm.
One approach to improve the algorithm is to encode outlier source symbols,
i.e., $|x_i|>\gamma$, explicitly using $\approx \log(|x_i|/\gamma)$ bits,
perhaps using a universal code for integers~\cite{Elias1975universal}. This encoder
would reduce the distortion caused by outliers, thus allowing
to use a narrower interval, yielding a reduction in the alphabet size.
We leave the study of outlier processing for future work, and focus instead
on using an adaptive reproduction alphabet to improve the algorithm.

\section{Adaptive reproduction alphabet algorithm}
\label{sec:adaptive}

Our approach to overcome the disadvantages
of large alphabets (Section~\ref{sec:naive}) is inspired by the ground-breaking
work by Rose on the discrete nature of the reproduction alphabet for iid sources
when the Shannon lower bound is not tight~\cite{Rose94}. In
many cases of interest, a small reproduction alphabet
achieves the RD function of an analog source.
Indeed, at sufficiently low rates even a binary reproduction alphabet is 
sometimes optimal~\cite{MarcoNeuhoff2006}. We thus focus on an algorithm
that, while supporting the possibility that the reproduction alphabet must
be large, also supports a possible reduction of the alphabet size,
while allowing the actual reproduction levels to adapt to the input.

\subsection{Adaptive reproduction levels}

Following the approach of Yang and Zhang~\cite{Yang1999},
we map the input $\xn$ to a sequence $\zn$ over a finite
alphabet $\Z$, where the actual output $\yn$ is derived
via a scalar function $y_i=a(z_i)$. Ideally, the function $a(\cdot)$ should
minimize expected distortion. Because we focus on square error distortion,
the optimal $a^*(\cdot)$ is the conditional expectation~\cite{Yang1999},
\begin{equation}
\label{eq:def:a^*}
a^*(\alpha)
= E[x_i|z_i=\alpha],\ \forall\alpha\in\Z.
\end{equation}

The decoder does not know $\xn$, and cannot
compute $a^*(\cdot)$. Therefore, we encode the numerical value
$a^*(\alpha)$ for each $\alpha\in\Z$.
We allocate $\b\log(\log(n))$ bits to encode each
\begin{equation}
\label{eq:def:a^*_q}
a^*_q(\alpha) \triangleq \frac{\lceil a^*(\alpha)\Delta\rceil} {\Delta},
\end{equation}
where $a^*_q(\alpha)$ is a quantized version of $a^*(\alpha)$, and 
the quantizer resolution $\frac{1}{\Delta}$ 
depends on $\b$ and the width of the interval being quantized.
We observe that it might be advantageous to allocate
more bits to encode $a^*_q(\alpha)$ for symbols $\alpha\in\Z$ that appear
more times in $\zn$, but leave such optimizations for future work.
Nonetheless, if some $\alpha\in\Z$ does not appear in $\zn$, then there is
no need to encode its numerical value. We expend one flag bit per
character of $\Z$ to describe the {\em effective alphabet} 
$\Ze= \Ze(\zn)$, where $\Ze\subseteq\Z$ is
the subset of the reproduction alphabet $\Z$ that appears in $\zn$.
Because $|\Z| = |\overY| = 2\lceil\log(n)\rceil^2+1$,  only 
$O(\log^2(n))$ flag bits are needed. In fact, it suffices for the encoder to
describe the cardinality of $\Ze$ using $O(\log(\log(n)))$ bits,
which is insignificant. 

The energy function (\ref{eq:MCMC_energy}) must be
modified to support adaptive alphabets as follows,
\begin{equation}
\label{eq:def:energy_adaptive}
\varepsilon_a(\zn) \triangleq n[H_k(\zn)-\beta d_a(\xn,\zn)]
+ \b\log(\log(n)) |\Ze(\zn)|,
\end{equation}
where $\b\log(\log(n)) |\Ze|$
bits are used to encode the reproduction levels that appear in
the effective alphabet $\Ze$, 
$d_a(\xn,\zn)$ is distortion with the adaptive alphabet,
\begin{equation}
\label{eq:def:d_a}
d_a(\xn,\zn)=d_n(\xn,a^*_q(\zn))
= \frac{1}{n} \sum_{i=1}^n (x_i - a^*_q(z_i))^2,
\end{equation}
$a^*_q(\zn)$ is shorthand for the $n$-tuple obtained 
by applying $a^*_q$ to the components of $\zn$,
and $a^*_q(\cdot)$ is computed using (\ref{eq:def:a^*}) and
(\ref{eq:def:a^*_q}).
These definitions require to modify
the previous Gibbs sampler (\ref{eq:y_n_i}) as follows,
\begin{eqnarray}
& & f_s(z_i=a|z^{n\backslash i})
\nonumber \\
&=& 
\frac{ 1 } {
\sum_b \exp\left\{ -s\left[ n\Delta H_k(z^{i-1}bz_{i+1}^n,a)
- \beta \Delta d_a(b,a,\zn) +
\b\log(\log(n)) \Delta\Ze(b,a) \right] \right\} },
\label{eq:Gibbs_tilde}
\end{eqnarray}
where
\begin{equation}
\label{eq:def:Delta_da_z}
\Delta d_a(b,a,\zn) \triangleq
n\left[ d_a(\xn,z^{i-1}bz_{i+1}^n) - d_a(\xn,z^{i-1}az_{i+1}^n) \right]
\end{equation}
is the change in distortion using the adaptive
alphabet (\ref{eq:def:d_a}),
and $\Delta\Ze(b,a)$ is the change in the size of the
effective alphabet when $z_i=a$ is replaced by $b$.
Alternately, the optimization routine can loop over different
alphabet sizes $|\Z|$
without accounting for $|\Z|$ in the energy (\ref{eq:def:energy_adaptive});
this latter approach was used in our simulations (Section~\ref{sec:numerical}).

The crux of the matter is that if a reduced alphabet yields
similar distortion results without increasing the coding length, then the
modified energy function (\ref{eq:def:energy_adaptive}) induces 
a smaller effective $\Ze$.
Motivated by the theoretical results by Rose~\cite{Rose94} and our numerical
results (Section~\ref{sec:numerical}), for many analog sources of
practical interest a small alphabet offers good and in many cases optimum
RD performance.
In such cases, the adaptive alphabet algorithm is advantageous.

Even if the entire alphabet is used, i.e., $\Ze=\Z$,
then the location of the reproduction levels is optimized via $a^*(\cdot)$
in lieu of the uniform quantization used in $\overY$
(\ref{eq:def:overlineY}).
Consequently, if we allow the adaptive alphabet algorithm to
use $\Z$ with the same cardinality of $\overY$
as in Algorithm~1, then the RD performance can only improve.

We now state formally that the adaptive alphabet algorithm achieves the RD
function asymptotically without prior knowledge of the source statistics.
As before, our result relies on the existence of a universal constant $c_2$ such that for
$c<c_2$ the transition probabilities between the $|\Z|^n$ possible outputs
are sufficiently large.

\begin{THEO}
\label{th:MCMC_adaptive}
Consider square error distortion (\ref{eq:def:d_n}),
let $X$ be a finite variance stationary and ergodic source with RD function
$R(X,D)$, use $\Z$ with cardinality $|\Z| = 2 \lceil\log(n)\rceil^2+1$
and sufficiently small $c<c_2$ in Algorithm~2, and let $a^*_q(\znr)$ be the MCMC approximation to $\xn$ after $r$
super-iterations. Then the length of context tree weighting (CTW)~\cite{Willems1995CTW} 
applied to $\znr$ converges as follows,
\[
\lim_{n\rightarrow\infty}
\lim_{r\rightarrow\infty}
E\left[ \frac{1}{n}|CTW(\znr)|-\beta d_a(\xn,\znr)\right]
\overset{n\rightarrow\infty}{\longrightarrow}
\min_{D \geq 0} [R(X,D)-\beta D].
\]
\end{THEO}

The formal proof appears in \ref{ap:MCMC_adaptive}. The key point
is that adaptive reproduction levels offer pointwise
improvement over the data-independent reproduction alphabet $\overY$
from Section~\ref{sec:naive}, per the same alphabet size. 

\subsection{Fast computation}

An important contribution by Jalali and Weissman~\cite{Jalali2008}
was to show how to compute
$\Delta H_k(y^{i-1}by_{i+1}^n,a)$ and $\Delta d(b,a,x_i)$ rapidly.
Without this computational contribution, the encoder would be impractical.
The adaptive algorithm updates $\Delta H_k(z^{i-1}bz_{i+1}^n,a)$
in an analogous manner. However, whereas 
$\Delta d(b,a,x_i)=(b-x_i)^2-(a-x_i)^2$ 
is trivial to compute for the  data-independent 
reproduction alphabet $\overY$ (\ref{eq:def:overlineY}), in our case
$\Delta d_a(b,a,\zn)$ (\ref{eq:def:Delta_da_z}) requires to re-compute
$d_a(\cdot,\cdot)$, which depends on $a^*_q(\cdot)$. Unfortunately,
modifying a single location in $\zn$ may change the distortion for numerous
symbols.

We now show how to compute
$\Delta d_a(b,a,\zn)$ rapidly for the adaptive reproduction alphabet 
algorithm. To do so, we evaluate $d_a(\xn,\zn)$,
\begin{eqnarray}
d_a(\xn,\zn)
&=&
\frac{1}{n} \sum_{i=1}^n d(x_i,a^*_q(z_i))
\label{eqn:d_a:step1} \\
&=&
\frac{1}{n} \sum_{\alpha\in\Z}
\sum_{\{i:\ z_i=\alpha\}}
\left(x_i-a^*_q(\alpha)\right)^2
\label{eqn:d_a:step2} \\
&=&
\frac{1}{n} \sum_{\alpha\in\Z}
\left\{
\sum_{\{i:\ z_i=\alpha\}}
\left[(x_i)^2\right]
- 2 a^*_q(\alpha)
\left[x_i\right]
+ (a^*_q(\alpha))^2
\left[1\right]
\right\},
\label{eqn:d_a:step3}
\end{eqnarray}
where (\ref{eqn:d_a:step1}) uses the definitions of $d_n(\cdot,\cdot)$
and $d_a(\cdot,\cdot)$ in (\ref{eq:def:d_n}) and (\ref{eq:def:d_a}), respectively,
and (\ref{eqn:d_a:step2}) partitions $z_i,\ i\in\{1,\ldots,n\}$, into the
different symbols $\alpha\in\Z$
and invokes the definition of square error distortion.
Combining (\ref{eq:def:a^*}) and (\ref{eq:def:a^*_q}),
\begin{equation}
a^*_q(\alpha)
=
\frac{\lceil E[x_i|z_i=\alpha] \Delta\rceil} {\Delta}
=
\frac{\left \lceil
\frac{ \sum_{\{i:\ z_i=\alpha\}} [x_i] }{ \sum_{\{i:\ z_i=\alpha\}} [1] }
\Delta \right\rceil} {\Delta}.
\label{eqn:a*_q_details}
\end{equation}
We see that (\ref{eqn:d_a:step3}) and (\ref{eqn:a*_q_details})
rely extensively on
\begin{equation}
\label{eq:def:X_alpha_k}
X_{\alpha}^m \triangleq \sum_{\{i:\ z_i=\alpha\}} \left[ (x_i)^m \right], \quad m\in\{0,1,2\}, \alpha \in \Z,
\end{equation}
the $m$'th moments of the portion of $x$ where $z_i=\alpha$.
In each iteration of the algorithm, a single $z_i$ may change from $\alpha$
to $\alpha'$. Consequently, we modify $X_{\alpha}^m$ and $X_{\alpha'}^m$,
$m\in\{0,1,2\}$, by adding and subtracting powers of $x_i$. Given these
updated values, the computation of $\Delta d_a(b,a,\zn)$ is rapid,
as before.
Pseudo-code for the adaptive alphabet Algorithm~2 appears below.

\fbox{\parbox{\algorithmwidth}{
\textsc{\underline{Algorithm~2}: Lossy encoder with adaptive reproduction alphabet} \\
\textsc{Input:} $\xn\in\mathbb{R}^n$, $\Z$, $\beta$, $c$, $r$ \\
\textsc{Output:} bit-stream \\
\textsc{Procedure:}
\begin{enumerate}
\vspace*{-3mm}
\item Initialize $z$ by quantizing $x$ // {\em can quantize with data-independent $\overY$}
\vspace*{-3mm}
\item Initialize $m_k(\cdot,\cdot)$ and other data structures using $z$
\vspace*{-3mm}
\item {\bf for} $t=1$ to $r$ {\bf do} {\em super-iteration}
\vspace*{-3mm}
\item \hspace*{5mm} $s \leftarrow c\log(t)$ for some $c>0$
// {\em inverse temperature}
\vspace*{-3mm}
\item \hspace*{5mm} Draw permutation of numbers $\{1,\ldots,n\}$ at random
\vspace*{-3mm}
\item \hspace*{5mm} {\bf for} $t'=1$ to $n$ {\bf do} 
\vspace*{-3mm}
\item \hspace*{10mm} Let~$i$ be component $t'$ in permutation
\vspace*{-3mm}
\item \hspace*{10mm} {\bf for} all $\alpha$ in $\Z$ {\bf do} // {\em evaluate possible changes to $z_i$ }
\vspace*{-3mm}
\item \hspace*{15mm} Compute $\Delta d_a(b,a,\zn)$ via
(\ref{eq:def:Delta_da_z}), (\ref{eqn:d_a:step3}), (\ref{eqn:a*_q_details}), and (\ref{eq:def:X_alpha_k})
\label{algo2:line_Delta_d}
\vspace*{-3mm}
\item \hspace*{15mm} Compute $f_s(z_i=\alpha|z^{n\backslash i})$ given in (\ref{eq:Gibbs_tilde})
// {\em modified Gibbs distribution}
\vspace*{-3mm}
\item \hspace*{10mm} Generate new $z_i$ using $f_s(z_i=\cdot|z^{n\backslash i})$ // {\em Gibbs sampling}
\vspace*{-3mm}
\item \hspace*{10mm} Update $m_k(\cdot,\cdot)[\cdot]$ and $X_{z_i}^m$, $m\in\{0,1,2\}$ // {\em previous and new $z_i$}
\vspace*{-3mm}
\item Encode effective alphabet $\Ze$ 
\vspace*{-3mm}
\item Encode $a^*_q(\alpha)$ using $\b \log(\log(n)) |\Ze|$ bits
\vspace*{-3mm}
\item Apply CTW to $\zn$ 
\end{enumerate}
}}

As for the data-independent reproduction alphabet case, Algorithm~2 requires 
$O(nk|\Z|)$ time to compute $\Delta H_k(z^{i-1}bz_{i+1}^n,a)$.
Utilizing the computational techniques specified above, $\Delta d_a(b,a,\zn)$ can
be computed in constant time per inner loop of each iteration 
(Line~\ref{algo2:line_Delta_d}), which requires
$O(n|\Z|)=O(n|\overY|)$ computation per super-iteration.
We see that computing $\Delta H_k(z^{i-1}bz_{i+1}^n,a)$
should require more time than computing $\Delta d_a(b,a,\zn)$;
this was confirmed in our implementation.

We have also noticed empirically that Algorithm~2 often 
comes quite close to optimum RD performance after a few 
dozen super-iterations, resulting in reasonable overall computational demands.
Additionally, in practice the effective alphabet $\Ze$ is often modest.
CTW~\cite{Willems1995CTW} converges to the empirical entropy as long as
$k_n=\log(n)/\log(|\Ze|)-\Omega_n(1)$, and for finite $n$ a smaller alphabet $|\Ze|$ allows CTW to converge
to the empirical entropy for larger context depths~$k_n$.
Therefore, Algorithm~2 can
optimize over deeper context trees, leading to improved compression and faster
convergence to the RD function.

The decoder of the adaptive reproduction alphabet Algorithm~2 resembles the
decoder in~\cite{Jalali2008}.
First, the bit-stream generated by CTW is decompressed
to reconstruct $\zn$. The actual real-valued reproduction sequence is
obtained by mapping from $\zn$ to $\yn$ via the adaptive quantizer $a^*_q(\alpha)$,
since the mapping $a^*_q$ has been described to the decoder.

\section{Numerical results}
\label{sec:numerical}

To demonstrate the potential of our approach, we implemented
the adaptive alphabet Algorithm~2 in Matlab. Results for
Laplace and autoregressive sources are provided.

{\bf Implementation details}:\
We ran Algorithm~2 for sequences of length $n=1.5\cdot 10^4$ using
$r=50$ super-iterations, and $k\approx\frac{1}{2}\log_{|\Z|}(n)$.
We also found two heuristics to be useful.
First, for each individual compression problem and RD slope $\beta$
the specific temperature evolution $s=O(\log(t))$ may vary.
Therefore, for each point we ran four temperature evolution
sequences and allowed each one to improve over the energy
$\varepsilon_a(\zn)$ computed with previous evolution sequences.
Second, using a good starting point helps  Algorithm~2 converge.
Therefore, we began running low rate problems with small~$\beta$, 
and each solution was used as a starting point for the next 
larger~$\beta$. 

Below we plot results averaging over 10 simulations.
Each plot compares the MCMC approach to 
entropy coding (ECSQ), results by Yang and Zhang~\cite{Yang1999},
and the RD function.

{\bf Laplace source}:\
We first evaluated an iid Laplace source with pdf 
$f(x)=\frac{1}{2}e^{-|x|}$ such that 
$E[X]=0$ and $\mbox{var}(X)=2$.
For this source, entropy coding performs rather well.
However, Yang and Zhang~\cite{Yang1999}
give better RD performance (Figure~\ref{fig:MCMC_Laplace}).
Algorithm~2 improves further over the deterministic minimization by Yang and
Zhang~\cite{Yang1999}, which requires availability of a training sequence.
Their algorithm can be used in a universal setting by partitioning
the input into blocks, resulting in a performance loss of 0.2--0.3
dB~\cite{Yang1999}. 

Relying on the mapping approach of Rose~\cite{Rose94}, it can be shown
that for low-to-medium rates a small odd number of reproduction levels
suffices to approach the fundamental RD limit.
We have observed that the optimal mapping $a^*(\alpha)$
is similar to the reproduction alphabet computed by
the mapping approach of Rose~\cite{Rose94}. This similarity suggests that applying
Algorithm~1 to the ``correct" finite alphabet would not
improve results by much.

\begin{figure*}[t]
\begin{center}
\includegraphics[width=140mm]{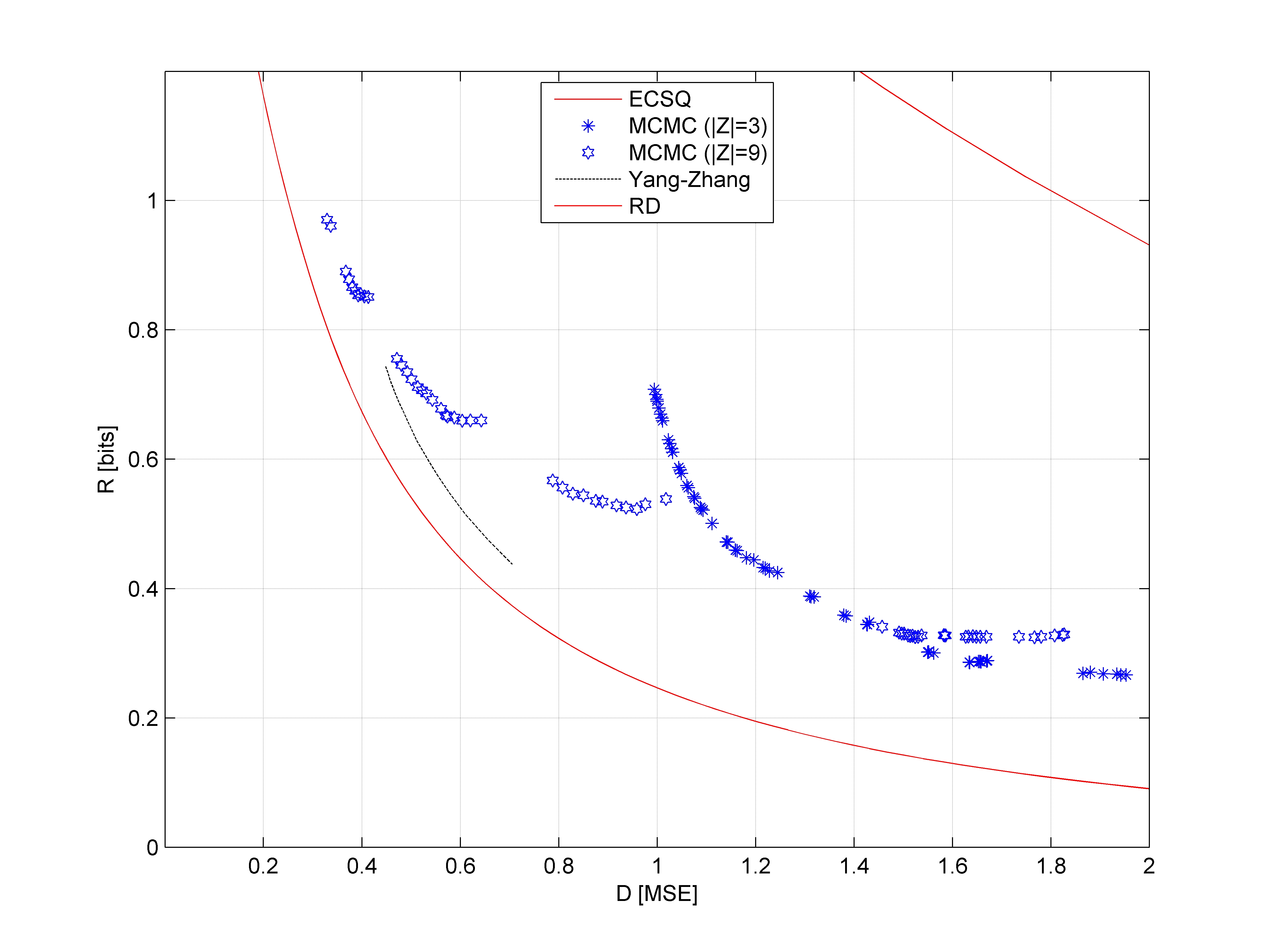}
\end{center}
\vspace*{-8mm}
\caption{
{\small\sl {\bf AR source}:\ Comparison of
entropy coding (ECSQ),
average rate and distortion of
Algorithm~2 (MCMC) over 10 simulations,
results by Yang and Zhang~\cite{Yang1999},
and the RD function.
($n=1.5\cdot 10^4$,  $|\Z|$$\in\{3,9\}$, $r=50$,
$k\approx\frac{1}{2}\log_{|\Z|}(n)$.)}
\label{fig:MCMC_AR}
}
\vspace*{-5mm}
\end{figure*}

{\bf Autoregressive source}:\
Figure~\ref{fig:MCMC_AR} illustrates the RD performance of the different
algorithms for an {\em autoregressive} (AR) source, where
\[
x_n=\rho x_{n-1} + w_n,
\]
$\rho=0.9$, and the innovation sequence
$w_n \sim \cal{N}$$(0,1)$ is zero mean unit norm iid Gaussian.

Entropy coding (ECSQ) is not well suited for non-iid sources;
vector quantization~\cite{GershoGray1993}, the deterministic
minimization algorithm by Yang and Zhang~\cite{Yang1999},
and MCMC  can be used instead.
Note that ECSQ appears in the upper right hand
side of the figure; its RD performance is poor.

We plotted the RD performance of Algorithm~2 using a small
reproduction alphabet ($|\Z|=3$) and a moderately sized one ($|\mbox{$\Z|$}=9$).
At low rates, the smaller alphabet offers better RD performance;
as the rate is increased, larger alphabets quantize the source more precisely.
Although Algorithm~2 does not compress as well as
Yang and Zhang (they describe the AR source as very challenging),
recall that Algorithm~2 is universal.

\section{Discussion}
\label{sec:discussion}

In this paper, we extended the MCMC simulated annealing approach of Jalali
and Weissman~\cite{Jalali2008} to analog
sources. We described two lossy compression algorithms that asymptotically
achieve the RD function universally for stationary ergodic
continuous amplitude sources. The first algorithm relies on a
data-independent reproduction alphabet that samples a wider interval with
finer resolution as the input length is increased. However, the large alphabet
slows down the convergence to the RD function, and is an impediment in practice.
Our second algorithm therefore uses a (potentially smaller) adaptive reproduction alphabet.
Not only is the adaptive algorithm theoretically motivated for iid sources
by the discrete nature of the reproduction alphabet when the Shannon
lower bound is not tight~\cite{Rose94}, but our numerical results
suggest that even for non-iid sources it works well. Additionally, 
the smaller alphabet accelerates the computation.

{\bf Applications}:\
In applications such as image compression~\cite{xiong97sf,Lopresto1997}, video
compression~\cite{Wiegand2003}, and speech coding~\cite{Makhoul85,buzo1980speech,sabin1984product},
our algorithms can process a vector of real-valued numbers whose statistics are
either completely unknown, or perhaps only known approximately.
The algorithms will then iterate over the data
until some reasonable RD performance is attained.

As an example, consider image coding. The EQ coder~\cite{Lopresto1997}
processes each sub-band of wavelets sequentially, going from low-frequency
sub-bands and proceeding toward high-frequency sub-bands.
The EQ coder classifies wavelet coefficients in each sub-band based on
the magnitudes of parent coefficients, relying on the insight that the
magnitudes of children coefficients are correlated with the
magnitudes of parents~\cite{Mallat1999book}. In a similar manner,
our algorithms can utilize the parent coefficients as contextual information.

\renewcommand\thesection{Appendix \Alph{section}}
\setcounter{section}{0}
\section{Proof of Theorem~\ref{th:achieve_fixed}}
\label{ap:achieve_fixed}

{\bf Continuous codebook}:\
We begin by constructing a continuous amplitude RD codebook~\cite{Berger71}.
Given the slope $\beta$ of the RD function $R(X,D)$, there exists an optimal rate
$R(X,\beta)$ and distortion level $D(X,\beta)$.
For any $\epsilon_1>0$, fix the rate $R=R(X,\beta)+\epsilon_1$.
The achievable RD coding theorem~\cite{Cover91,Berger71} demonstrates for the
source $X$ that in the limit of large $n$ there exist codebooks whose rates are smaller 
than $R$ and whose expected per symbol distortions are  less than $D(X,\beta)$.
We choose such a codebook $\C$ comprised of at most $2^{Rn}$ codewords,
each of length $n$. 
The encoder maps $\xn$ to the nearest
codeword $c_j$ in $\C$ and transmits its index~$j$.
The decoder then maps index~$j$ to $c_j$.

{\bf Quantized codebook}:\
Now define a quantized codebook $\overC$ such that
$\overcij$, the $i$'th entry of the $j$'th codeword of
$\overC$, is generated by rounding $c_{ij}$,
the $i$'th entry of the $j$'th codeword of $\C$,
to the closest value in $\overY$.
(Recall that $-\gamma$ and $\gamma$ are the smallest and largest
values in $\overY$, respectively.)
Using $\overC$, the encoder and decoder are identical, except
that $\overcij$ is used instead of $c_{ij}$.
The quantized codebook $\overC$ requires the same
rate as before.\footnote{
By quantizing $c_j$ to $\overcj$, different $c_j$ could yield identical
codewords in $\overC$; this would allow to reduce the rate.}
However, the distortion provided by the 
quantized codebook $\overC$ is different.

{\bf Distortion of quantized codebook}:\
To analyze the change in distortion, we consider three cases.
In the first case, the original codebook value is an outlier whereas
the signal value $x_i$ is not, i.e., $|c_{ij}| > \gamma$ and $|x_i|\leq\gamma$.
The truncation of $|c_{ij}|$ to $\gamma$ reduces the distortion,
\[
d(x_i,\overcij)
= (x_i-\overcij)^2 
< (x_i-c_{ij})^2
= d(x_i,c_{ij}).
\]
The second case occurs when the original codebook and signal values
are both outliers, i.e., $|x_i|,|c_{ij}| > \gamma$.
As $n\rightarrow\infty$, the amount of variance beyond the increasing
$\gamma=\lceil\log(n)\rceil$ vanishes,
$E[(x_i\cdot 1_{\{ |x_i|>\gamma(n) \}})^2]\overset{n\rightarrow\infty}{\longrightarrow}0$,
because the source $X$ has finite variance and $\gamma$ increases
(\ref{eq:def:overlineY}).
Therefore, for any $\delta_1>0$ there exists $N_1$ such that for all $n>N_1$
the increase in expected distortion $d_n(\xn,\yn)$ (\ref{eq:def:d_n})
due to truncation of outliers is smaller than $\delta_1$.
The third case occurs for $|c_{ij}| \leq \gamma$,
where rounding changes the square error from
$(x_i-c_{ij})^2$ to $(x_i-\overcij)^2$, and the distortion changes by
\begin{eqnarray*}
(x_i-\overcij)^2 - (x_i-c_{ij})^2
&=&
(\overcij)^2 - (c_{ij})^2 + 2x_i(c_{ij}-\overcij) \\
&=&
(c_{ij}-\overcij )
(2x_i - \overcij - c_{ij} ) \\
&=&
(c_{ij}-\overcij )
\left[ 2(x_i - c_{ij}) + (c_{ij} - \overcij) \right].
\end{eqnarray*}
Because $|c_{ij}-\overcij|\leq\frac{1}{2\gamma}$ (\ref{eq:def:overlineY}),
the change in distortion is upper bounded as follows,
\[
|(x_i-\overcij)^2 - (x_i-c_{ij})^2|
\leq
\frac{|x_i - c_{ij}|}{\gamma} + \frac{1}{4\gamma^2}.
\]
We now define sets of indices that relate to the three cases,
\begin{align*}
\I_1 &\triangleq& \{i:\ i\in\{1,\ldots,n\}, |c_{ij}| > \gamma, |x_i|\leq\gamma \}, \\
\I_2 &\triangleq& \{i:\ i\in\{1,\ldots,n\}, |x_i|,|c_{ij}| > \gamma \}, \\
\I_3 &\triangleq& \{i:\ i\in\{1,\ldots,n\}, |c_{ij}| \leq \gamma \}.
\end{align*}
Summarizing over all $i\in\{1,\ldots,n\}$,
\begin{eqnarray}
E\left[ nd_n(\xn,\overcj) \right]
&=&
E\left[ \sum_{i=1}^n (x_i-\overcij)^2 \right]
\nonumber \\
&=&
E\left[ \left[ \sum_{i\in \I_1} (x_i-\overcij)^2 \right] +
\left[ \sum_{i\in \I_2} (x_i-\overcij)^2 \right] +
\left[ \sum_{i\in \I_3} (x_i-\overcij)^2 \right] \right]
\nonumber \\
&\leq &
E\left[ \sum_{i\in \I_1} (x_i-c_{ij})^2 \right] +
E\left[ \sum_{i\in \I_2} (x_i-c_{ij})^2 + n\delta_1 \right]
\nonumber \\
& &
+E\left[ \sum_{i\in \I_3} (x_i-c_{ij})^2 + \frac{|x_i-c_{ij}|}{\gamma}
+\frac{1}{4\gamma^2} \right]
\label{eqn:th:achieve_fixed:D1} \\
&\leq&
nE\left[d_n(\xn,c_j)\right] 
+ n\delta_1 + \frac{E\left[\|\xn-c_j\|_1\right]}{\gamma}
+\frac{n}{4\gamma^2},
\label{eqn:th:achieve_fixed:D2}
\end{eqnarray}
where $c_j$ and $\overcj$ are the $j$'th codewords of $\C$
and $\overC$, respectively,
$\|\cdot\|_1$ denotes the $\ell_1$ norm,
the inequality in (\ref{eqn:th:achieve_fixed:D1}) relies on
the changes in distortion in the three different cases,
and the inequality in (\ref{eqn:th:achieve_fixed:D2}) is due to the $\gamma$
terms related to $\I_3$ that do not appear for $\I_1$
and $\I_2$.
Because $E[d_n(\xn,c_j)]\leq D$
and $\|\xn-c_j\|_1 \leq n\sqrt{d_n(\xn,c_j)}$, we have
via Jensen's inequality that $E[\|\xn-c_j\|_1] \leq n\sqrt{D}$.
Therefore,
\begin{equation}
\label{eq:ap:th2_D1}
E[d_n(\xn,\overcj)]
<
D + \delta_1 + \frac{\sqrt{D}}{\gamma} +\frac{1}{4\gamma^2}
=
D + \delta_1 + \frac{\sqrt{D}}{\lceil\log(n)\rceil} + \frac{1}{4\lceil\log(n)\rceil^2}.
\end{equation}
Because $\gamma=\lceil\log(n)\rceil$ increases with $n$, 
\begin{equation}
\label{eq:ap:th2_D2}
E[d(\xn,\overcj)] \leq D + 2\delta_1.
\end{equation}
Therefore, the quantized codebook $\overC$ approaches the RD function
asymptotically for the continuous amplitude source $X$.

{\bf Lossless compression using CTW}:\
Having demonstrated that there exists a codebook based on the finite alphabet
$\overC$ that asymptotically
achieves the RD function, we need to prove that the RD performance of
$\overC$ can be approached by compressing  $\xhatn$ losslessly using 
CTW~\cite{Willems1995CTW}.
The remainder of the proof borrows from the
prior art on lossy compression of finite sources~\cite{Yang1997,Jalali2008}.
Owing to the linearity of expectation,
\begin{equation}
\label{eq:th:achieve_fixed:expectation}
E\left[ \frac{1}{n}|CTW(\xhatn)|-\beta d(\xn,\xhatn)\right]
=
E\left[ \frac{1}{n}|CTW(\xhatn)| - H_k(\xhatn) \right]
+ E\left[ H_k(\xhatn) - \beta d(\xn,\xhatn) \right].
\end{equation}
Recall that $k=k_n=o(\log(n))$, and so for any $\epsilon_2>0$ there exists
$N_2$ such that for all $n>N_2$ CTW converges to the empirical
entropy~\cite{Willems1995CTW},
\begin{equation}
\label{eq:th:achieve_fixed:CTW_converge}
E\left[ \frac{1}{n}|CTW(\xhatn)| - H_k(\xhatn) \right] < \epsilon_2,
\end{equation}
as long as $k_n=\log(n)/\log(|\overY|)-\Omega_n(1)$.
Jalali and Weissman~\cite{Jalali2008} invoke Gray et al.~\cite{GrayNO1975}
to prove that for any $\delta_2>0$ and $\epsilon_3>0$ there exists a process
$\widetilde{X}$ that is jointly stationary and ergodic with $X$ such that
\begin{eqnarray}
E\left[ H_k(\xhatn) - \beta d(\xn,\xhatn) \right]
&\leq&
E\left[ H_k(\widetilde{\xn}) - \beta d(\xn,\widetilde{\xn}) \right]
\label{eqn:th:achieve_fixed:Gray1} \\
&\leq&
H(\widetilde{X_0}|\widetilde{X_{-k}^{-1}})+\epsilon_3
- E\left[ \beta d(\xn,\widetilde{\xn}) \right]
\label{eqn:th:achieve_fixed:Gray2} \\
&\leq&
R(X,D)+\epsilon_4 + \epsilon_3 -\beta (D(\beta)+\delta_2),
\label{eqn:th:achieve_fixed:Gray3}
\end{eqnarray}
where (\ref{eqn:th:achieve_fixed:Gray1}) relies on the definition
of $\xhatn$ (\ref{eq:def:xhat}),
(\ref{eqn:th:achieve_fixed:Gray2}) is explained by observing that
$H_k(\widetilde{\xn})$ converges to $H(\widetilde{X_0}|\widetilde{X_{-k}^{-1}})$
with probability one as $k_n$ is increased,
and (\ref{eqn:th:achieve_fixed:Gray3}) uses properties of $\widetilde{X}$,
i.e.,
$H(\widetilde{X_0}|\widetilde{X_{-k}^{-1}}) \leq R(X,D) + \epsilon_4 $ and
$E\left[ \beta d(\xn,\widetilde{\xn}) \right] \leq D(\beta)+\delta_2$.
Note also that $R(X,D)$ relies implicitly on $\beta$, and is identical
to the $R(\beta)$ mentioned earlier.
We complete the proof by combining
(\ref{eq:ap:th2_D2}),
(\ref{eq:th:achieve_fixed:expectation}),
(\ref{eq:th:achieve_fixed:CTW_converge}),
(\ref{eqn:th:achieve_fixed:Gray3}),
and the arbitrariness of $\delta_1$, $\delta_2$, $\epsilon_2$, $\epsilon_3$,
and $\epsilon_4$ 
\hfill$\Box$

\section{Proof of Theorem~\ref{th:MCMC_fixed}}
\label{ap:MCMC_fixed}

The proof is similar to that by Jalali and
Weissman~\cite[Appendix~B]{Jalali2008}, and we only outline the
arguments here. While the algorithm is running, $\ynr$
takes one of $|\overY|^n$ possible values. These values
are modeled as states of a Markov chain.
There is a sufficiently positive probability
to transition between any two states, because 
$c<c_1$ and each super-iteration
of Algorithm~1 processes all $n$ locations of $\yn$.
If the temperature is reduced slowly enough, then the distribution
of $\ynr$ converges toward the stationary distribution of the Markov chain.
The proof is completed by noting that minimal-energy states occupy all
the probabilistic mass of the stationary distribution. Therefore,
$\ynr$ converges in distribution to the set of minimal energy solutions,
and we enjoy the same RD performance as in Theorem~\ref{th:achieve_fixed}.
\hfill$\Box$

\section{Proof of Theorem~\ref{th:MCMC_adaptive}}
\label{ap:MCMC_adaptive}

The proof is similar to the proofs of
Theorems~\ref{th:achieve_fixed} and~\ref{th:MCMC_fixed}.
Consider the sequence $\zhatn$ with globally minimal
modified energy~(\ref{eq:def:energy_adaptive}),
\begin{equation}
\label{eq:def:xa_hat}
\zhatn \triangleq
\arg\min_{w^n \in \Z^n} \varepsilon_a(w^n).
\end{equation}
We first employ arguments from~\ref{ap:achieve_fixed} to prove
that $\zhatn$ achieves the RD function asymptotically.
Next, we prove that simulated annealing~\cite{Geman1984}
converges to the globally optimal solution asymptotically.

{\bf Achievable for global minimum}:\
Recall that the adaptive algorithm uses $\Z$ with cardinality
$|\Z|=|\overY|$.
Consider~\ref{ap:achieve_fixed}, which proves that the globally
optimal data-independent reproduction alphabet solution $\xhatn$ achieves the
RD function asymptotically. Because $|\Z|=|\overY|$,
there exists a one to one mapping from $\overY$ to
$\Z$, and $\xhatn$ is mapped to some $\widetilde{\zn}$.
The optimal $a^*(\cdot)$ may reduce the distortion,
\[
d_n(\xn,a^*(\zhatn)) \leq d_n(\xn,\xhatn).
\]
Although the quantized version $a^*_q(\cdot)$ may increase the
distortion, i.e.,
\[
d_a(\xn,\zhatn)
= d_n(\xn,a^*_q(\zhatn))
\geq d_n(\xn,a^*(\zhatn)),
\]
allocating $\b\log(\log(n))$ bits to encode each $a^*_q(\alpha)$
is sufficient to guarantee that the quantization error is
smaller than $\frac{1}{\gamma}$ (see the proof of
Theorem~\ref{th:achieve_fixed} in~\ref{ap:achieve_fixed}).
Our previous derivations
(\ref{eqn:th:achieve_fixed:D2}), (\ref{eq:ap:th2_D1}), (\ref{eq:ap:th2_D2})
show that for any $\delta>0$ the overall distortion $d_a(\xn,\zhatn)$
becomes $\delta$-close to $D(\gamma)$ as $n$ is increased.
We conclude from the definitions of energy (\ref{eq:MCMC_energy})
and modified adaptive energy (\ref{eq:def:energy_adaptive}) that
\[
\varepsilon_a(\zhatn) \leq
\varepsilon(\xhatn) + n\beta\delta + \b \log(\log(n))  |\overY| .
\]
Because $|\overY|=O(\log^2(n))$, the last
term due to encoding the quantized $a^*_q(\cdot)$ vanishes relative
to $n\beta\delta$.
Taking $\delta$ as small as we want enables to approach
$\min_{D \geq 0} [R(X,D)-\beta D]$ as closely as needed,
\[
\lim_{n\rightarrow\infty}\sup
E\left[ \frac{1}{n}|CTW(\zhatn)|-\beta d_a(\xn,\zhatn)\right] \leq
\min_{D \geq 0} [R(X,D)-\beta D].
\]
Invoking the converse result of Yang et al.~\cite{Yang1999,Yang1997},
\[
E\left[ \frac{1}{n}|CTW(\zhatn)|-\beta d_a(\xn,\zhatn)\right]
\overset{n\rightarrow\infty}{\longrightarrow}
\min_{D \geq 0} [R(X,D)-\beta D].
\]

{\bf Simulated annealing}:\
The proof is similar to that in~\ref{ap:MCMC_fixed}.
The only noteworthy point is
that for each $\zn$ the quantized $a^*_q(\cdot)$ is a deterministic
function of $\zn$. Therefore, the simulated annealing can be
posed as a Markov chain over $|\Z|^n$ states, where convergence in
distribution to the set of minimal energy solutions is obtained by
recognizing that for $c<c_2$ there is a sufficiently positive probability to transition between
any two states.
\hfill$\Box$

\section*{Acknowledgments}

This work was supported by Israel Science Foundation grant number 2013433. 
The first author thanks the generous hospitality of the 
Electrical Engineering Department at the Technion,
Israel, where much of this work was performed.
We also thank Shirin Jalali for sharing her software
implementations of the algorithms in~\cite{Jalali2008}, and Ken Rose for
enlightening discussions on the discrete nature of the codebook~\cite{Rose94}.

{\small
\bibliographystyle{IEEEtran}
\bibliography{cites}

\begin{thebibliography}{10}
\providecommand{\url}[1]{#1}
\csname url@samestyle\endcsname
\providecommand{\newblock}{\relax}
\providecommand{\bibinfo}[2]{#2}
\providecommand{\BIBentrySTDinterwordspacing}{\spaceskip=0pt\relax}
\providecommand{\BIBentryALTinterwordstretchfactor}{4}
\providecommand{\BIBentryALTinterwordspacing}{\spaceskip=\fontdimen2\font plus
\BIBentryALTinterwordstretchfactor\fontdimen3\font minus
  \fontdimen4\font\relax}
\providecommand{\BIBforeignlanguage}[2]{{%
\expandafter\ifx\csname l@#1\endcsname\relax
\typeout{** WARNING: IEEEtran.bst: No hyphenation pattern has been}%
\typeout{** loaded for the language `#1'. Using the pattern for}%
\typeout{** the default language instead.}%
\else
\language=\csname l@#1\endcsname
\fi
#2}}
\providecommand{\BIBdecl}{\relax}
\BIBdecl

\bibitem{BaronWeissmanDCC2010}
D.~Baron and T.~Weissman, ``An {MCMC} approach to lossy compression of
  continuous sources,'' in \emph{Proc. Data Compression Conf. (DCC)}, Mar.
  2010, pp. 40--48.

\bibitem{xiong97sf}
Z.~Xiong, K.~Ramchandran, and M.~T. Orchard, ``Space-frequency quantization for
  wavelet image coding,'' \emph{IEEE Trans. Image Process.}, vol.~6, no.~5, pp.
  677--693, May 1997.

\bibitem{Lopresto1997}
S.~M. Lopresto, K.~Ramchandran, and M.~T. Orchard, ``Image coding based on
  mixture modeling of wavelet coefficients and a fast estimation-quantization
  framework,'' in \emph{Proc. Data Compression Conf. (DCC)}, Mar. 1997, pp.
  221--230.

\bibitem{Wiegand2003}
T.~Wiegand, G.~J. Sullivan, G.~Bjontegaard, and A.~Luthra, ``Overview of the
  {H.264/AVC} video coding standard,'' \emph{IEEE Trans. Circuits Syst. Video
  Technol.}, vol.~13, no.~7, pp. 560--576, Jul. 2003.

\bibitem{Makhoul85}
J.~Makhoul, S.~Roucos, and H.~Gish, ``Vector quantization in speech coding,''
  \emph{Proc. IEEE}, vol.~73, no.~11, pp. 1551--1588, Nov. 1985.

\bibitem{buzo1980speech}
A.~Buzo, A.~Gray~Jr, R.~Gray, and J.~Markel, ``{Speech coding based upon vector
  quantization},'' \emph{IEEE Trans. Acoustics, Speech and Signal Process.},
  vol.~28, no.~5, pp. 562--574, 1980.

\bibitem{sabin1984product}
M.~Sabin and R.~Gray, ``{Product code vector quantizers for waveform and voice
  coding},'' \emph{IEEE Trans. Acoustics, Speech and Signal Process.}, vol.~32,
  no.~3, pp. 474--488, 1984.

\bibitem{FarvardinModestino84}
N.~Farvardin and J.~W. Modestino, ``Optimum quantizer performance for a class
  of non-{G}aussian memoryless sources,'' \emph{IEEE Trans. Inf. Theory},
  vol.~30, no.~3, pp. 485--496, May 1984.

\bibitem{Lloyd82}
S.~P. Lloyd, ``Least squares quantization in {PCM},'' \emph{IEEE Trans. Inf.
  Theory}, vol.~28, no.~2, pp. 129--137, Mar. 1982.

\bibitem{Max60}
J.~Max, ``Quantization for minimum distortion,'' \emph{IRE Trans. Inf. Theory},
  vol.~6, no.~1, pp. 7--12, Mar. 1960.

\bibitem{Huffman52}
D.~A. Huffman, ``A method for the construction of minimum-redundancy codes,''
  \emph{Proc. Inst. Radio Eng.}, vol.~9, no.~40, pp. 1098--1101, Sep. 1952.

\bibitem{RissanenLangdon1981}
J.~Rissanen and J.~G.~Langdon, ``Universal modeling and coding,'' \emph{IEEE
  Trans. Inf. Theory}, vol.~27, no.~1, pp. 12--23, Jan. 1981.

\bibitem{Cover91}
T.~M. Cover and J.~A. Thomas, \emph{Elements of Information Theory}.\hskip 1em
  plus 0.5em minus 0.4em\relax Wiley-Interscience, 1991.

\bibitem{Berger71}
T.~Berger, \emph{Rate distortion theory; a mathematical basis for data
  compression}.\hskip 1em plus 0.5em minus 0.4em\relax Prentice-Hall Englewood
  Cliffs, NJ, 1971.

\bibitem{chou1989entropy}
P.~Chou, T.~Lookabaugh, and R.~Gray, ``{Entropy-constrained vector
  quantization},'' \emph{IEEE Trans. Acoustics, Speech and Signal Process.},
  vol.~37, no.~1, pp. 31--42, 1989.

\bibitem{riskin1991greedy}
E.~Riskin and R.~Gray, ``{A greedy tree growing algorithm for the design of
  variable rate vector quantizers [image compression]},'' \emph{IEEE Trans.
  Signal Process.}, vol.~39, no.~11, pp. 2500--2507, 1991.

\bibitem{GershoGray1993}
A.~Gersho and R.~M. Gray, \emph{{Vector quantization and signal
  compression}}.\hskip 1em plus 0.5em minus 0.4em\relax Kluwer, 1993.

\bibitem{GioranKontoyiannis2009}
C.~Gioran and I.~Kontoyiannis, ``Lossy compression in near-linear time via
  efficient random codebooks and databases,'' \emph{CoRR}, vol. abs/0904.3340,
  2009.

\bibitem{GVW2008b}
A.~Gupta, S.~Verd{\'u}, and T.~Weissman, ``Linear-time near-optimal lossy
  compression,'' in \emph{Proc. Int. Symp. Inf. Theory (ISIT2008)}, Jul. 2008.

\bibitem{Kontoyiannis1999}
I.~Kontoyiannis, ``{An implementable lossy version of the {L}empel-{Z}iv
  algorithm - {P}art {I}: {O}ptimality for memoryless sources},'' \emph{IEEE
  Trans. Inf. Theory}, vol.~45, no.~7, pp. 2293--2305, Nov. 1999.

\bibitem{Zamir2001}
R.~Zamir and K.~Rose, ``{Natural type selection in adaptive lossy
  compression},'' \emph{IEEE Trans. Inf. Theory}, vol.~47, no.~1, pp. 99--111,
  Jan. 2001.

\bibitem{Yang1997}
E.~Yang, Z.~Zhang, and T.~Berger, ``{Fixed-slope universal lossy data
  compression},'' \emph{IEEE Trans. Inf. Theory}, vol.~43, no.~5, pp.
  1465--1476, Sep. 1997.

\bibitem{Jalali2008}
S.~Jalali and T.~Weissman, ``Rate-distortion via {M}arkov chain {M}onte
  {C}arlo,'' in \emph{Proc. Int. Symp. Inf. Theory (ISIT2008)}, Jul. 2008, pp.
  852--856.

\bibitem{Hussami2009}
N.~Hussami, S.~B. Korada, and R.~L. Urbanke, ``Polar codes for channel and
  source coding,'' \emph{CoRR}, vol. abs/0901.2370, 2009.

\bibitem{LinderZamir94}
T.~Linder and R.~Zamir, ``On the asymptotic tightness of the {S}hannon lower
  bound,'' \emph{IEEE Trans. Inf. Theory}, vol.~40, no.~6, pp. 2026--2031, Nov.
  1994.

\bibitem{GLZ99}
A.~Gy{\"o}rgy, T.~Linder, and K.~Zeger, ``On the rate-distortion function of
  random vectors and stationary sources with mixed distributions,'' \emph{IEEE
  Trans. Inf. Theory}, vol.~45, no.~6, pp. 2110--2115, Sep. 1999.

\bibitem{RosenthalBinia88}
H.~Rosenthal and J.~Binia, ``On the epsilon entropy of mixed random
  variables,'' \emph{IEEE Trans. Inf. Theory}, vol.~34, no.~5, pp. 1110--1114,
  Sep. 1988.

\bibitem{WeidmannV2008}
C.~Weidmann and M.~Vetterli, ``Rate distortion behavior of sparse sources,''
  2008, submitted.

\bibitem{WeidmannV99}
------, ``Rate-distortion analysis of spike processes,'' in \emph{Proc. Data
  Compression Conf. (DCC)}, Mar. 1999, pp. 82--91.

\bibitem{CWO2002}
R.~Castro, M.~B. Wakin, and M.~Orchard, ``On the problem of simultaneous
  encoding of magnitude and location,'' in \emph{Asilomar Conf. Signals, Syst.,
  Comput.}, 2002.

\bibitem{ChangSparseRD2009}
C.~Chang, ``On the rate distortion function of {B}ernoulli {G}aussian
  sequences,'' \emph{CoRR}, vol. abs/0901.3820, 2009.

\bibitem{MarcoNeuhoff2006}
D.~Marco and D.~L. Neuhoff, ``Low-resolution scalar quantization for {G}aussian
  sources and squared error,'' \emph{IEEE Trans. Inf. Theory}, vol.~52, no.~4,
  pp. 1689--1697, Apr. 2006.

\bibitem{Yang1999}
E.~Yang and Z.~Zhang, ``{Variable-rate trellis source encoding},'' \emph{IEEE
  Trans. Inf. Theory}, vol.~45, no.~2, pp. 586--608, Mar. 1999.

\bibitem{Geman1984}
S.~Geman and D.~Geman, ``Stochastic relaxation, {G}ibbs distributions, and the
  {B}ayesian restoration of images,'' \emph{IEEE Trans. Pattern Anal. Mach.
  Intell.}, vol.~6, pp. 721--741, Nov. 1984.

\bibitem{Rose94}
K.~Rose, ``A mapping approach to rate-distortion computation and analysis,''
  \emph{IEEE Trans. Inf. Theory}, vol.~40, no.~6, pp. 1939--1952, Nov. 1994.

\bibitem{Willems1995CTW}
F.~M.~J. Willems, Y.~Shtarkov, and T.~J. Tjalkens, ``The context tree weighting
  method: {B}asic properties,'' \emph{IEEE Trans. Inf. Theory}, vol.~41, no.~3,
  pp. 653--664, May 1995.

\bibitem{LZ78}
J.~Ziv and A.~Lempel, ``Compression of individual sequences via variable-rate
  coding,'' \emph{IEEE Trans. Inf. Theory}, vol.~24, no.~5, pp. 530--536, Sep.
  1978.

\bibitem{Elias1975universal}
P.~Elias, ``Universal codeword sets and representations of the integers,''
  \emph{IEEE Trans. Inf. Theory}, vol.~21, no.~2, pp. 194--203, Mar. 1975.

\bibitem{Mallat1999book}
S.~Mallat, \emph{{A wavelet tour of signal processing}}.\hskip 1em plus 0.5em
  minus 0.4em\relax Academic Press, 1999.

\bibitem{GrayNO1975}
R.~Gray, D.~Neuhoff, and J.~Omura, ``{Process definitions of distortion-rate
  functions and source coding theorems},'' \emph{Trans. Inf. Theory}, vol.~21,
  no.~5, pp. 524--532, Sep. 1975.

\end{thebibliography}
}
\end{document}